# Electrical Properties of Crystalline $Ba_{0.5}Sr_{0.5}TiO_3$ Thin Films

Irzaman[1], H. Darmasetiawan[1], M. Nur Indro[1], S. G. Sukaryo[1], M Hikam[2], Na Peng Bo[2], M. Barmawi[3]

[1]Department of Physics, FMIPA IPB, Jl. Raya Pajajaran Bogor, Indonesia
Telp. : 62-(0251)-326166, fax. : 62-(0251)-326166,e-mail : fisika1@fmipa.ipb.ac.id
[2]Department of Physics, FMIPA University Indonesia, Kampus Depok, Indonesia
Telp. : 62-(021)-9137323, fax. : 62-(021) 7863441,e-mail : hikam@makara.cso.ui.ac.id
[3]Department of Physics, FMIPA ITB, Jl. Ganesha 10 Bandung, Indonesia
Telp. 62-(022) 251 1848, fax. : 62-(022) 251 1848

## ABSTRACT

Thin $Ba_{0.5}Sr_{0.5}TiO_3$ (BST) films on *p*-type Si (100) using The Chemical Solution Deposition (CSD) method. X-ray diffraction (XRD), Scanning electron microscopy (SEM), C-V meter analysis measurement were employed to characterise the films. The growth condition to obtain enough quality epitaxial of $Ba_{0.5}Sr_{0.5}TiO_3$ carried out by spin coating at 3000 rpm for 30 seconds, and then annealing at 900°C for 15 hours. The structure and crystallinity of thin films were investigated by XRD preffered orientation (100), (010); (110), (111), (200), surface analysis by SEM magnification x10000 thin films were heterogen and thickness film 1100 nm; electrical characterization $Ba_{0.5}Sr_{0.5}TiO_3$ at MFS (Metal Ferroelectric Semiconductor) structure in room temperature (300 K) by carried out capacitance flat band ($C_{FB}$) for frequency 10 KHz and 100 KHz were 206 pF and 187 pF, dielectric constant ($\varepsilon$) for frequency 10 KHz and 100 KHz were 132.67 and 117.22, dielectric loss (tg$\delta$) minimum for frequency 10 KHz and 100 KHz were 5.37 % and 6.43 % respectively.

## I.    INTRODUCTION

Thin $Ba_{0.5}Sr_{0.5}TiO_3$ is a well known dielectric material and has been attractive for the applications such as capacitors and high density dynamic random access memory (DRAM) due to its high dielectric constant and high capacity of charge storage [1,2].

$Ba_xSr_{1-x}TiO_3$ films can be formed by various methods, such as Chemical Solution Deposition (CSD)[1,2,3,], Metal Organic Chemical Vapor Deposition (MOCVD)[4,5,6], RF Sputtering [7,8,9,10,11,12] and Pulsed Laser Ablation Deposition (PLAD) [13].

CSD Method is of particular interest because of its good control of stoichiometry, ease of fabrication and low temperature synthesis. It is relatively new and requires a greater understanding to optimize film quality. Crystallization mechanisms in CSD-derived thin films are different from phenomena associated with vapor phase epitaxy. It was reported that CSD derived thermodynamically stable.

## II. METHODOLOGY

Procedural $Ba_{0.5}Sr_{0.5}TiO_3$ thin films were fabricated by CSD methods using 0.1916 gram Barium acetic $[Ba(CH_3COO)_2, 99\%]$ + 0.1543 gram Strontium acetic $[Sr(CH_3COO)_2, 99\%]$ + 0.4264 gram Titanium isopropoxide $[Ti(C_{12}O_4H_{28}), 99.999\%]$ as precursor in 5 ml 2-methoxyethanol $[H_3COOCH_2CH_2OH, 99.9\%]$ was used as solvent was introduced under mixing by Ultrasonic Model Branson 2210 at 1 hour. A clear liquid resulted. After 20 minutes of standing at room temperature, this solution acquired a milky appearance. 0.18 ml of acetic acid was introduced into the solution and mixed. The cloudy appearance disappeared. It contained equivalent 0.3 M $Ba_{0.5}Sr_{0.5}TiO_3$. After 2 hours of aging, the solution described above was applied on a silicon (100) substrate with measure 12 mm x 12 mm with a spin coating 3000 rpm for 30 seconds. The coated was then baked at 900°C for 15 hours on air atmosphere in Furnace Model Nabertherm Type 27. The structure of the films was analyzed by x-ray diffraction (XRD). The XRD patterns were recorded on a Diano type 2100E diffractometer using CuKα radiation at 30 KV and 30 mA (900 watt). The surface morphology and thickness of the films was examined by SEM JEOL type JSM-35C. The electrical capacitance – voltage (C-V) characterization $Ba_{0.5}Sr_{0.5}TiO_3$ by LCZ meter Model 2343 NF and Keithley 617 current source. Step by research to explain within flow diagram in Figure 1.

Primary goal of this experiment is dielectric films at MFS (Metal Ferroelectric Semiconductor) structure. Of particular interest is the value of capacitance the flat-band condition ($C_{FB}$) and Debye Length ($L_D$) is Eq. (1) and Eq (2) namely : [14]

$$C_{FB} = \frac{A}{\frac{d}{\varepsilon_0 \varepsilon_F} + \frac{L_D}{\varepsilon_0 \varepsilon_s}}, \qquad (1)$$

$$L_D = \sqrt{\frac{\varepsilon_0 \varepsilon_s \phi_t}{qN_A}}, \qquad (2)$$

where $C_{FB}$ = capacitor flat-band (farad), $L_D$ = Debye Length (m), A = Aluminum contact area = 2.5 x $10^{-7}$ m$^2$; d = thickness films = 1,1 x $10^{-6}$ m, $\varepsilon_0$ = permitivity relative vacuum = 8.854 x $10^{-12}$ C$^2$/N m$^2$, $\varepsilon_F$ = dielectric constant $Ba_{0.5}Sr_{0.5}TiO_3$ film , $\varepsilon_s$ = dielectric constant silicon = 11,9, $\phi_t$ = kT/q the thermal voltage (volt) at room temperature = 2,59 x $10^{-2}$ volt, k = Boltzmann's constant = 1.38 x $10^{-23}$, T = absolute temperature = 300 K, q = electron charge = 1.6 x $10^{-19}$ coulomb, $N_A$ = acceptor impurity density *p*-type silicon (100) substrate = 2 x $10^{22}$ m$^{-3}$.

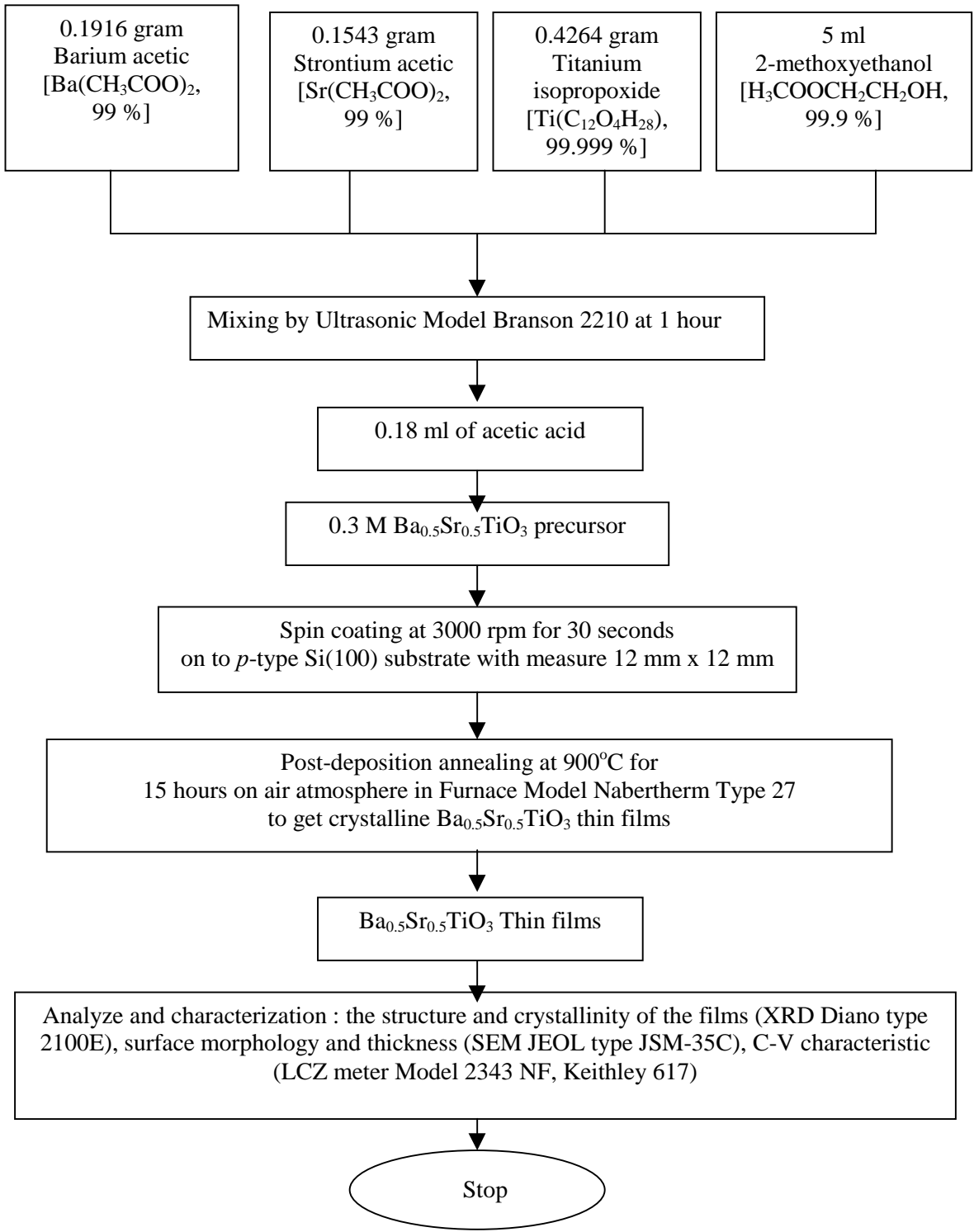

Figure 1. Flow diagram of Step by Step Research

## II. RESULT AND DISCUSSION

Figure 2 shows the structure and crystallinity of $Ba_{0.5}Sr_{0.5}TiO_3$ thin films were investigated by XRD, at 3000 rpm preferred orientation $Ba_{0.5}Sr_{0.5}TiO_3$ (100), (010); (110), (111), (200) appeared.

Figure 3 shows the morphology film by SEM magnification x10000 surface thin films were heterogen because not yet optimized parameter deposition and cross section carried out thickness film (d) was 1100 nm. Figure 4 shows C-V characteristics of $Ba_{0.5}Sr_{0.5}TiO_3$ films and dielectric loss at various source frequency in room temperature (300 K) by carried out capacitance flat band ($C_{FB}$) for frequency 10 KHz and 100 KHz were 206 pF and 187 pF, exactly dielectric loss (tgδ) minimum for frequency 10 KHz and 100 KHz were 5.37 % and 6.43 % respectively. Using formula (1) and (2) carried out dielectric film ($\varepsilon_F$) at frequency 10 KHz and 100 KHz were 132.67 and 117.22 respectively.

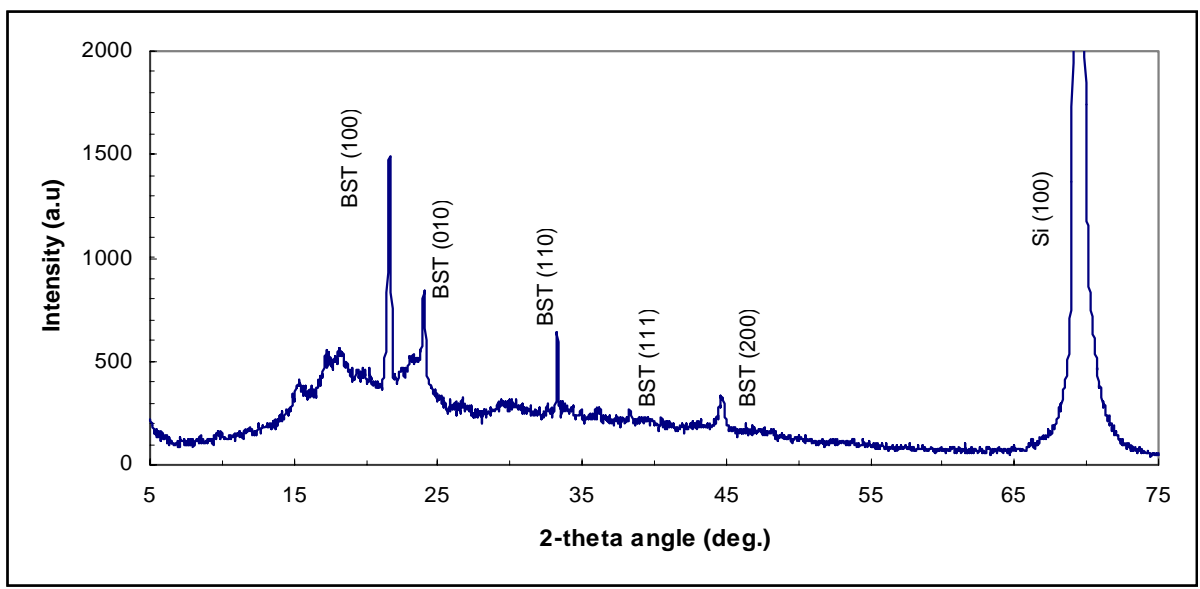

Figure 2. The structure and crystallinity by XRD of $Ba_{0.5}Sr_{0.5}TiO_3$ (BST) thin films on Si (100) substrate parameter deposition ω = 3000 rpm for 30 seconds, temperature annealing 900°C for 15 hours.

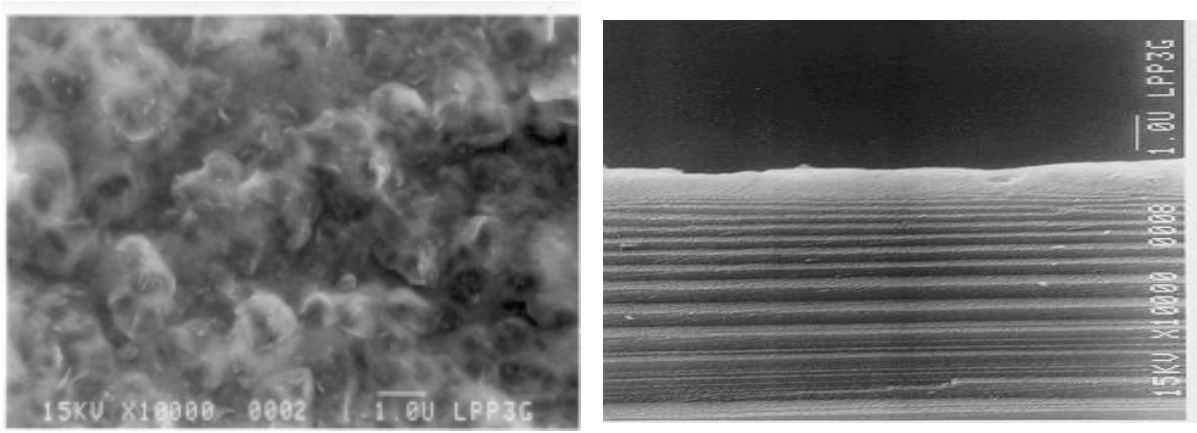

(a) (b)

Figure 3. (a) The morphology $Ba_{0.5}Sr_{0.5}TiO_3$ (BST) film by SEM magnification x10000 for ω = 3000 rpm, (b) Cross section film thickness (d) was 1100 nm.

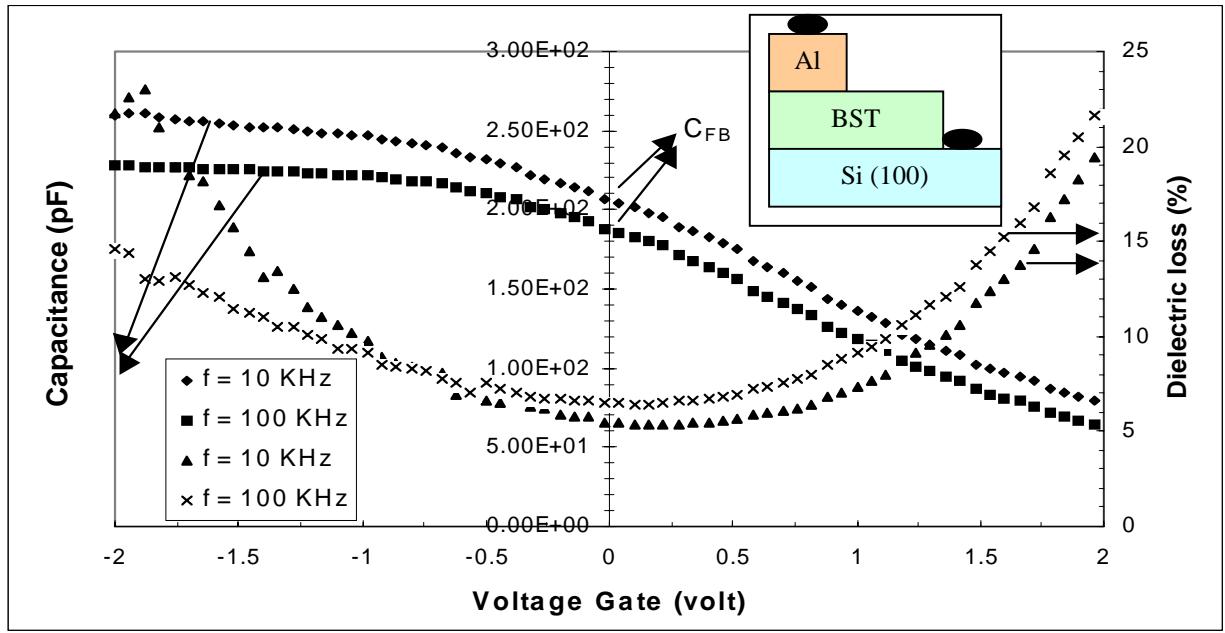

Figure 4. MFS (Metal Ferroelectric Semiconductor) C-V curves and dielectric loss (tgδ) curves measured at room temperature (300 K) with frequency a parameter, the Si substrate is *p*-type.

Figure 4 shows C-V characteristics of $Ba_{0.5}Sr_{0.5}TiO_3$ thin films and dielectric loss at various source frequency in room temperature (300 K) by carried out capacitance flat band ($C_{FB}$) at frequency 10 KHz and 100 KHz were 206 pF and 187 pF, exactly dielectric loss (tg δ) minimum at frequency 10 KHz and 100 KHz were 5.37 % and 6.43 % respectively. Using formula (1) and (2) carried out dielectric film ($\varepsilon_F$) at frequency 10 KHz and 100 KHz were 132.67 and 117.22 respectively.

IV. CONCLUSIONS

The growth condition to obtain enough quality epitaxial of $Ba_{0.5}Sr_{0.5}TiO_3$ on *p*-type Si (100) substrate The growth condition to obtain enough quality epitaxial of $Ba_{0.5}Sr_{0.5}TiO_3$ carried out by spin coating at 3000 rpm for 30 seconds, and then annealing at 900°C for 15 hours. The structure and crystallinity of thin films were investigated by XRD preferred orientation (100), (010); (110), (111), (200), surface analysis by SEM (at 10000x magnification) thin films were heterogen and the thickness film 1100 nm.

Electrical properties Ba$_{0.5}$Sr$_{0.5}$TiO$_3$ at MFS (Metal Ferroelectric Semiconductor) structure at room temperature (300 K) by carried out capacitance flat band (C$_{FB}$) at frequency 10 KHz and 100 KHz were 206 pF and 187 pF, dielectric constant ($\varepsilon$) at frequency 10 KHz and 100 KHz were 132.67 and 117.22, dielectric loss (tg $\delta$) minimum at frequency 10 KHz and 100 KHz were 5.37 % and 6.43 % respectively.

ACKNOWLEDGMENT

This work was supported by DCRG URGE Project, The Ministry of National Education, The Republic of Indonesia, under contract No. 011/DCRG/URGE/2000.